# Basis Set Extrapolation from the Vanishing Counterpoise Correction Condition

Vladimir Fishman, Emmanouil Semidalas, and Jan M. L. Martin*

**ABSTRACT:** Basis set extrapolations are typically rationalized either from analytical arguments involving the partial-wave or principal expansions of the correlation energy in helium-like systems or from fitting extrapolation parameters to reference energetics for a small(ish) training set. Seeking to avoid both, we explore a third alternative: extracting extrapolation parameters from the requirement that the BSSE (basis set superposition error) should vanish at the complete basis set limit. We find this to be a viable approach provided that the underlying basis sets are not too small and reasonably well balanced. For basis sets not augmented by diffuse functions, BSSE minimization and energy fitting yield quite similar parameters.

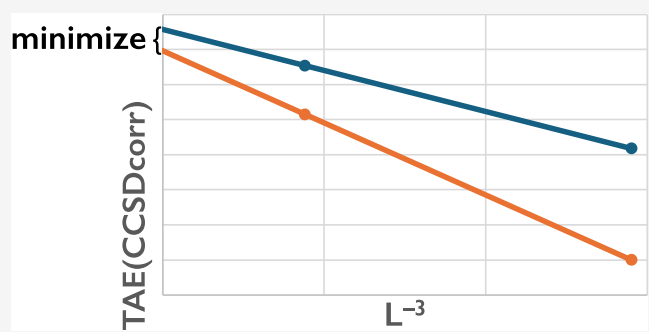

## INTRODUCTION

Despite great recent progress in density functional theory, wave function ab initio methods such as coupled cluster theory can still routinely exceed the accuracy of the best DFT calculations by an order of magnitude, provided they are close enough to the one-particle basis set limit.

For atom-centered orbital basis sets, basis set convergence of the correlation energy is excruciatingly slow. Schwartz[1,2] showed in the early 1960s that for the second-order correlation energy of helium-like atoms, the contributions of successive angular momenta (the "partial waves") converge as

$$E_l^{(2)} = A/(l + 1/2)^4 + B/(l + 1/2)^6 + ... \quad (1)$$

Then if the basis set is truncated at angular momentum $L$, the total residual error is

$$E_\infty^{(2)} - E_L^{(2)} = \sum_{l=L+1}^{\infty} A/(l + 1/2)^4 + B/(l + 1/2)^6 + ... \quad (2)$$

$$= \frac{A\psi^{(3)}(L + 3/2)}{6} + \frac{B\psi^{(5)}(L + 3/2)}{120} + ... \quad (3)$$

where $\psi$ is the polygamma function. For large $L$, this function can be approximated by the asymptotic series

$$\psi^{(3)}(L + 3/2) = 2/(L + 1)^3 + O(L^5) \quad (4)$$

and

$$\psi^{(5)}(L + 3/2) = 24/(L + 1)^5 + O(L^7) \quad (5)$$

Hill[3] generalized this result to configuration interaction, while Kutzelnigg and Morgan[4] derived a general leading $L^{-3}$ dependence for singlet-coupled, and $L^{-5}$ for triplet-coupled, pair correlation energies. The latter authors also showed that in the presence of explicit $R_{12}$ terms in the basis set, convergence will asymptotically be accelerated to $L^{-7}$.

A similar leading $\propto L^{-3}$ dependence is obtained from two different sets of considerations. Carroll, Silverstone, and Metzger in 1979 showed[5] that the basis set convergence in the principal expansion asymptotically converges as $\delta E_{nlm} = -A/\left(n - \frac{1}{2}\right)^6$. For a given principal quantum number $n$, however, the angular quantum number $l$ runs from 0 to $n - 1$, and the magnetic quantum number $m$ from $-l$ to $+l$. This leads to $\sum_{l=0}^{n-1}(2l + 1) = n^2$ approximately equal contributions, and hence an overall $\propto n^{-4}$ leading dependence. Summing over all missing shells, from $n_{max}+1$ to infinity, again leads us to a leading inverse-cubic $\propto n^{-3}$ dependence of eq 4.

Later, Petersson and co-workers[6–9] considered the convergence of the correlation energy in a natural orbital expansion, and found it to converge as $\propto N^{-1}$ (with $N$ the number of natural orbitals retained) for opposite-spin correlation, and $\propto N^{5/3}$ for same-spin correlation. As the number of natural orbitals in a basis set series such as the correlation consistent[10] cc-pVnZ or atomic natural orbital[11] ANO-n will converge with the cardinal number n as

$$N = (n+1)(n+3/2)(n+2)/3$$
$$= \left(n+\frac{3}{2}\right)^3 - \frac{1}{12}\left(n+\frac{3}{2}\right) \quad (6)$$

we once again recover an inverse-cubic dependence. (See also Klopper.[12] For an illustration with natural orbitals in neon atom, see Figure 1 in the present work. The natural orbitals there were obtained from the *spdfghi* primitives in the cc-pV10Z basis set of Feller et al.[13])

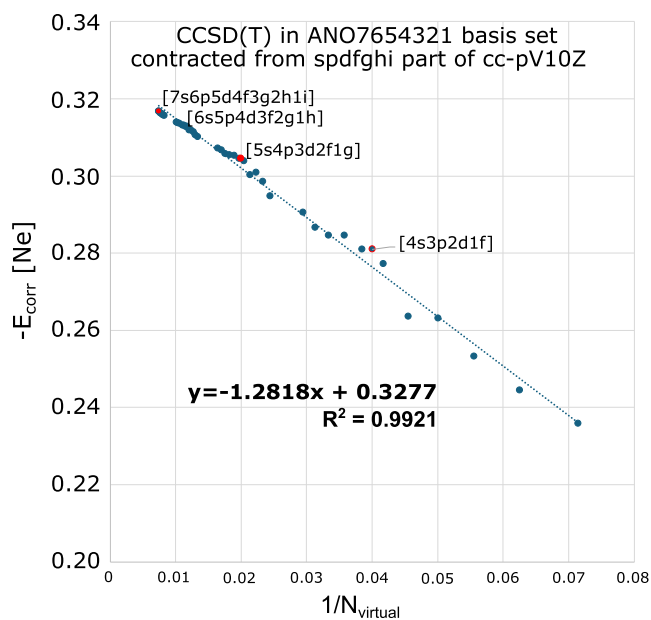

**Figure 1.** Convergence of CCSD(T) correlation energy of neon atom as a function of the number of natural orbitals included. Natural orbitals obtained from the *spdfghi* part of the cc-pV10Z basis set.

Applying such extrapolation formulas to the basis set convergence in *molecules* entails a major leap of faith. In the mid-nineties, Helgaker and co-workers[14,15] and Martin[16] found that this works well enough in practice; Klopper[17] introduced the additional refinement that the correlation energy is partitioned between same-spin (strictly: "triplet-coupled pair") and opposite-spin (strictly speaking: "singlet-coupled pair") contributions, and that these contributions are extrapolated separately assuming $L^{-5}$ and $L^{-3}$ behavior, respectively. (The partitioning is not unique for open-shell systems: see ref 18)

Several variants have been introduced, such as those with variable exponents $\alpha$ of the form $E(L) = E_\infty + A/L^\alpha$ (e.g., ref 19), variable L-shift $E_\infty + A/(L + a)^3$ (Petersson[20,21]), variable cardinal numbers $X(L)$ for the basis sets (Varandas[22]), etc. As explained in ref 23, all of them can be related to the same linear two-point extrapolation of Schwenke[24]

$$E_\infty \approx E_L + A_L(E_L - E_{L-1}) \quad (7)$$

where we will refer to $A_L$ as a "Schwenke coefficient", which is specific to the level of theory and the basis set pair.

Further work by Schwenke[25] going up to $L = 12$ appears to indicate that after initial rapid convergence, a "diminishing returns" regime quickly sets in.

**Basis Set Superposition Error.** BSSE (basis set superposition error) results when an interaction energy between monomers A and B is evaluated in a finite basis as $E(AB) - E(A) - E(B)$, where A only carries the basis functions of monomer A, and likewise for B. If the basis set on A is far from the CBS (complete basis set) limit, the availability in the dimer of the additional basis functions from the other monomer leads to an artifactual stabilization of the dimer known as BSSE.

Particularly in calculations on noncovalent interactions, BSSE can rival the interaction energy itself unless well-saturated and balanced basis sets are used.

The classic remedy is the counterpoise method,[26] in which the monomer energies are effectively evaluated in the whole dimer basis set. BSSE can then be defined operationally as the difference between "raw" and corrected interaction energies.

$$\begin{aligned} \text{BSSE} &= E[A] + E[B] - E[A(B)] - E[B(A)] \\ &= E[AB] - E[A(B)] - E[B(A)] \\ &\quad - (E[AB] - E[A] - E[B]) \\ &= D_e[\text{raw}] - D_e[\text{CP}] \end{aligned} \quad (8)$$

At the complete basis set limit, BSSE should be zero — and hence if an extrapolation works correctly, then the "raw" and counterpoise answers should be the same. Discrepancies thus indicate either a flaw in the extrapolation formula, or inadequate basis sets, or both.

We now propose to invert this observation — by using the requirement that BSSE should be zero, or minimized, as a means of obtaining basis set extrapolations.

This has the advantage that it relies neither on the theoretical behavior for an idealized system, nor on fitting (possibly themselves flawed) reference interaction energies for some training dataset.

To the best of our knowledge, the concept of deriving a basis set extrapolation from the BSSE limiting condition has never been explored. However, the NASA Ames team, in the late 1980s, did advocate using a negative multiple of the calculated BSSE as a correction for basis set incompleteness (e.g., refs 27,28). Quoting Taylor[28]

Since BSSE is in some sense a measure of basis set incompleteness, one can contemplate increasing the bond energy by some fraction of the counterpoise correction to correct for this residual incompleteness, rather than decreasing it to correct the computed result for BSSE. This is a completely empirical approach, but we have found (for strong interactions) that in large basis sets (up to g functions, say) increasing the computed values by 150% of the calculated BSSE gives a good approximation to the best extrapolations to the basis set limit that we can perform from very large basis set studies.

## ■ COMPUTATIONAL DETAILS

All quantum chemical calculations were performed using either MOLPRO 2024.1[29] or Gaussian 16 rev. C.01[30] running on the CHEMFARM cluster of the Faculty of Chemistry at Weizmann.

Three basis set sequences were considered:

1. the nZaPa sequence ($n = 2-7$) of Ranasinghe and Petersson (RP)[21]
2. the augmented correlation consistent sequence of Dunning:

**Table 1. Schwenke Extrapolation Coefficients $A_L$ and RMS Deviations (kcal·mol$^{-1}$) from 24 Diatomics Dataset and from the Literature**

| | | Schwenke parameters $A_L$ | | | | | RMS(BSSE) or RMSD(TAE) | | | | TAE with BSSE parameter and vice versa | | | |
|---|---|---|---|---|---|---|---|---|---|---|---|---|---|---|
| | | {T,Q} | {Q,5} | {5,6} | {6,7} | | {T,Q} | {Q,5} | {5,6} | {6,7} | {T,Q} | {Q,5} | {5,6} | {6,7} |
| CCSD raw | V$n$Z+d | 0.759 | 0.924 | 1.162 | 1.391 | TAE RAW | 0.205 | 0.080 | 0.046 | 0.040 | 0.232 | 0.103 | 0.069 | 0.089 |
| CCSD CP | V$n$Z+d | 0.751 | 0.922 | 1.140 | 1.472 | TAE CP | 0.233 | 0.058 | 0.041 | 0.044 | 0.259 | 0.099 | 0.067 | 0.087 |
| CCSD BSSE | V$n$Z+d | 0.722 | 0.869 | 1.060 | 1.719 | BSSE | 0.076 | 0.056 | 0.029 | 0.033 | 0.085 | 0.060 | 0.034 | 0.040 |
| | Schwenke[24] | 0.700 | 0.900 | 1.238 | | | | | | | | | | |
| | Varandas[51] | 0.635 | 0.849 | 1.142 | | | | | | | | | | |
| CCSD raw | nZaPa | 0.705 | 0.887 | 1.120 | 1.452 | TAE RAW | 0.183 | 0.080 | 0.053 | 0.037 | 0.183 | 0.119 | 0.054 | 0.058 |
| CCSD CP | nZaPa | 0.710 | 0.869 | 1.130 | 1.499 | TAE CP | 0.235 | 0.094 | 0.043 | 0.035 | 0.236 | 0.130 | 0.044 | 0.057 |
| CCSD BSSE | nZaPa | 0.706 | 0.805 | 1.148 | 1.669 | BSSE | 0.116 | 0.048 | 0.020 | 0.008 | 0.116 | 0.057 | 0.020 | 0.016 |
| CCSD raw | haV$n$Z+d | 0.647 | 0.892 | 1.228 | 1.453 | TAE RAW | 0.128 | 0.095 | 0.041 | 0.041 | 0.356 | 0.095 | 0.070 | 0.075 |
| CCSD CP | haV$n$Z+d | 0.677 | 0.906 | 1.189 | 1.541 | TAE CP | 0.153 | 0.056 | 0.043 | 0.031 | 0.361 | 0.059 | 0.073 | 0.067 |
| CCSD BSSE | haV$n$Z+d | 0.774 | 0.892 | 1.071 | 1.799 | BSSE | 0.106 | 0.059 | 0.014 | 0.025 | 0.148 | 0.059 | 0.027 | 0.032 |
| | Varandas[51] | 0.665 | 0.912 | 1.295 | [1.592] | | | | | | | | | |
| | Schwenke[24] | 0.700 | 0.930 | 1.266 | [1.621] | | | | | | | | | |
| | ref 23 | N/A | 0.932 | 1.283 | 1.602 | | | | | | | | | |
| (T) RAW | V$n$Z+d | 0.755 | 0.834 | 1.090 | 1.469 | TAE RAW | 0.032 | 0.013 | 0.005 | 0.004 | 0.038 | 0.015 | 0.009 | 0.009 |
| (T) CP | V$n$Z+d | 0.764 | 0.833 | 1.110 | 1.411 | TAE CP | 0.038 | 0.010 | 0.004 | 0.003 | 0.043 | 0.012 | 0.008 | 0.009 |
| BSSE (T) | V$n$Z+d | 0.715 | 0.792 | 1.001 | 1.692 | BSSE | 0.015 | 0.007 | 0.002 | 0.003 | 0.016 | 0.007 | 0.003 | 0.003 |
| | Schwenke[24] | 0.695 | 0.741 | 1.102 | | | | | | | | | | |
| (T) RAW | nZaPa | 0.678 | 0.841 | 1.097 | 1.562 | TAE RAW | 0.020 | 0.007 | 0.003 | 0.001 | 0.055 | 0.013 | 0.005 | 0.003 |
| (T) CP | nZaPa | 0.703 | 0.829 | 1.109 | 1.583 | TAE CP | 0.022 | 0.007 | 0.003 | 0.001 | 0.055 | 0.013 | 0.005 | 0.003 |
| BSSE (T) | nZaPa | 0.562 | 0.908 | 1.043 | 1.466 | BSSE | 0.004 | 0.003 | 0.001 | 0.001 | 0.011 | 0.003 | 0.002 | 0.001 |
| RP,[21] eq 12 | nZaPa | 0.604 | 0.891 | 1.199 | 1.517 | | | | | | | | | |
| RP,[21] optimized | nZaPa | 0.600 | 0.849 | 1.164 | 1.580 | | | | | | | | | |
| (T) RAW | haV$n$Z+d | 0.758 | 0.823 | 1.166 | 1.558 | (T) RAW | 0.030 | 0.009 | 0.003 | 0.003 | 0.049 | 0.019 | 0.014 | 0.006 |
| (T) CP | haV$n$Z+d | 0.727 | 0.805 | 1.209 | 1.526 | (T) CP | 0.032 | 0.007 | 0.003 | 0.003 | 0.055 | 0.018 | 0.013 | 0.006 |
| BSSE (T) | haV$n$Z+d | 0.864 | 0.932 | 0.941 | 1.763 | BSSE (T) | 0.014 | 0.005 | 0.001 | 0.001 | 0.015 | 0.005 | 0.003 | 0.002 |
| | Schwenke | 0.700 | 0.810 | 1.248 | | | | | | | | | | |

- aug-cc-pV$n$Z for first row: ref 31
- aug-cc-pV(n+d)Z for second-row elements[32,33] (concerning why second-row elements in high oxidation states need tight 3d functions added, see ref 34 and references therein)
- cc-pV7Z hydrogen, aug-cc-pV7Z carbon through fluorine: refs 35,36
- sulfur aug-cc-pV(7+d)Z from ESI of ref 37 (see also ref 38)

3. the core−valence correlation versions[39,40] of the above, but used for valence correlation only. It has previously been shown[41] that this practice considerably reduces BSSE.

The CCSD(T)[42,43] electronic structure method was used throughout. For open-shell systems, we adopted the Watts-Gauss-Bartlett definition[43] of restricted open-shell CCSD(T).

The molecules considered in the present work were all taken from the W4-17 thermochemical benchmark.[44] Reference geometries given in its Supporting Information, each optimized at the CCSD(T)/cc-pV(Q+d)Z level, were used as-is without reoptimization.

Throughout the paper, notation like cc-pV{T,Q}Z refers to extrapolation, in the given example from cc-pVTZ and cc-pVQZ basis sets. The shorthands pVTZ+d, haVTZ+d, CVTZ, and haCVTZ refer, respectively, to cc-pV(T+d)Z, heavy-aug-cc-pV(T+d)Z, cc-pCVTZ, and heavy-aug-cc-pCVTZ. (The common practice of omitting diffuse functions on hydrogen, while placing them on more electronegative elements, goes by several names in the literature: aug′-cc-pV$n$Z by Del Bene,[45] heavy-aug-cc-pV$n$Z by Hobza,[46] and jul-cc-pV$n$Z in "calendar sets" notation.[47])

For BSSE evaluation in polyatomics, we exclusively use the SSFC (site−site function counterpoise) of Wells and Wilson,[48] as implemented in MOLPRO's scripting language by one of us. Operationally, SSFC entails evaluating all monomer energies in the full oligomer basis set: the unmodified procedure may be inefficient for large clusters (where some sort of screening is called for ref 49) but this is not an issue in small-molecule systems of the W4-17 type.

In principle, one could for each pair of basis sets and for each molecule $i$ evaluate the $A_{L,i}$ that would make the extrapolated BSSE vanish

$$A_{L,i} = \text{BSSE}_{L,i}/(\text{BSSE}_{L-1,i} - \text{BSSE}_L) \qquad (9)$$

and then take the average over all molecules in the test set $\overline{A_{L,i}} = \sum_i A_{L,i}/n$. However, a more solid approach would seem to be least-squares minimization with respect to $A_L$ of the aggregate BSSE over the test set.

Table 2. Schwenke Extrapolation Parameters and RMS Deviations (kcal·mol$^{-1}$) in the CCSD Correlation Component of the TAE for Various Basis Set Sequences[a]

| | | BSSE-fitted | | | TAE-fitted | | |
|---|---|---|---|---|---|---|---|
| | | TQ | Q5 | 56 | TQ | Q5 | 56 |
| **VnZ+d** | $A_L$ 24diatom | 0.721 | 0.874 | 1.063 | 0.748 | 0.896 | 1.094 |
| | $A_L$ W4-08 | 0.719 | 0.881 | 1.081 | 0.689 | 0.883 | 1.109 |
| | $A_L$ W4-11 | 0.725 | 0.845 | 1.076 | 0.678 | 0.902 | 1.120 |
| | $A_L$ W4-17most | 0.719 | 0.870 | 1.082 | 0.690 | 0.899 | 1.119 |
| | RMSD BSSE(TAE) | 0.103 | 0.079 | 0.046 | 0.176 | 0.114 | 0.048 |
| | RMSD TAE | 0.514 | 0.119 | 0.064 | 0.495 | 0.119 | 0.060 |
| | LR slope | 0.698 | 0.815 | 0.989 | 0.700 | 0.878 | 1.127 |
| | $2\sigma$(slope) | 0.024 | 0.046 | 0.063 | 0.046 | 0.028 | 0.034 |
| | LR intercept (kcal·mol$^{-1}$) | 0.041 | 0.049 | 0.029 | -0.058 | 0.010 | -0.015 |
| | $R^2$ | 0.973 | 0.932 | 0.914 | 0.908 | 0.976 | 0.979 |
| **aVnZ+d** | $A_L$ 24diatom | 0.774 | 0.893 | 1.084 | 0.634 | 0.855 | 1.133 |
| | $A_L$ W4-08 | 0.783 | 0.884 | 1.121 | 0.602 | 0.867 | 1.149 |
| | $A_L$ W4-11 | 0.820 | 0.840 | 1.145 | 0.590 | 0.907 | 1.166 |
| | $A_L$ W4-17most | 0.777 | 0.875 | 1.133 | 0.593 | 0.908 | 1.171 |
| | RMSD BSSE(TAE) | 0.163 | 0.086 | 0.022 | 0.282 | 0.087 | 0.023 |
| | RMSD TAE | 0.797 | 0.128 | 0.060 | 0.279 | 0.126 | 0.058 |
| | LR slope | 0.724 | 0.824 | 1.078 | 0.632 | 0.829 | 1.149 |
| | $2\sigma$(slope) | 0.048 | 0.062 | 0.038 | 0.028 | 0.036 | 0.044 |
| | LR intercept (kcal·mol$^{-1}$) | 0.088 | 0.038 | 0.011 | -0.139 | 0.059 | 0.000 |
| | $R^2$ | 0.908 | 0.882 | 0.972 | 0.955 | 0.957 | 0.967 |
| **nZaPa** | $A_L$ 24diatom | 0.730 | 0.798 | 1.148 | 0.693 | 0.856 | 1.045 |
| | $A_L$ W4-08 | 0.717 | 0.826 | 1.169 | 0.642 | 0.850 | 1.071 |
| | $A_L$ W4-11 | 0.720 | 0.820 | 1.139 | 0.635 | 0.863 | 1.103 |
| | $A_L$ W4-17most | 0.735 | 0.807 | 1.151 | 0.633 | 0.858 | 1.096 |
| | RMSD BSSE(TAE) | 0.141 | 0.064 | 0.031 | 0.181 | 0.066 | 0.039 |
| | RMSD TAE | 0.572 | 0.106 | 0.096 | 0.466 | 0.099 | 0.072 |
| | LR slope | 0.666 | 0.769 | 1.109 | 0.665 | 0.861 | 1.069 |
| | $2\sigma$(slope) | 0.055 | 0.017 | 0.028 | 0.036 | 0.035 | 0.019 |
| | LR intercept | 0.093 | 0.041 | 0.016 | -0.114 | -0.020 | 0.001 |
| | $R^2$ | 0.860 | 0.989 | 0.986 | 0.934 | 0.962 | 0.993 |
| **CVnZ** | $A_L$ 24diatom | 0.685 | 0.863 | 1.217 | 0.758 | 0.875 | 1.058 |
| | $A_L$ W4-08 | 0.686 | 0.883 | 1.217 | 0.699 | 0.857 | 1.078 |
| | $A_L$ W4-11 | 0.692 | 0.856 | 1.194 | 0.688 | 0.871 | 1.090 |
| | $A_L$ W4-17most | 0.686 | 0.872 | 1.194 | 0.676 | 0.866 | 1.090 |
| | RMSD BSSE(TAE) | 0.076 | 0.066 | 0.026 | 0.078 | 0.068 | 0.040 |
| | RMSD TAE | 0.511 | 0.113 | 0.119 | 0.507 | 0.103 | 0.059 |
| | LR slope | 0.661 | 0.825 | 1.181 | 0.708 | 0.854 | 1.100 |
| | $2\sigma$(slope) | 0.017 | 0.042 | 0.043 | 0.047 | 0.024 | 0.033 |
| | LR intercept | 0.046 | 0.039 | 0.010 | -0.044 | 0.007 | -0.018 |
| | $R^2$ | 0.984 | 0.943 | 0.969 | 0.906 | 0.982 | 0.979 |
| **haCVnZ** | $A_L$ 24diatom | 0.733 | 0.877 | 1.290 | 0.648 | 0.837 | 1.058 |
| | $A_L$ W4-08 | 0.736 | 0.886 | 1.309 | 0.622 | 0.843 | 1.091 |
| | $A_L$ W4-11 | 0.770 | 0.871 | 1.289 | 0.612 | 0.866 | 1.112 |
| | $A_L$ W4-17most | 0.741 | 0.877 | 1.288 | 0.616 | 0.866 | 1.112 |
| | RMSD BSSE(TAE) | 0.127 | 0.031 | 0.032 | 0.177 | 0.036 | 0.048 |
| | RMSD TAE | 0.560 | 0.101 | 0.138 | 0.281 | 0.081 | 0.055 |
| | LR slope | 0.669 | 0.861 | 1.277 | 0.647 | 0.824 | 1.109 |
| | $2\sigma$(slope) | 0.040 | 0.028 | 0.076 | 0.029 | 0.023 | 0.040 |
| | LR intercept (kcal·mol$^{-1}$) | 0.087 | 0.013 | 0.006 | -0.121 | 0.032 | -0.012 |
| | $R^2$ | 0.922 | 0.977 | 0.923 | 0.956 | 0.982 | 0.970 |

[a]"LR" refers to linear regression with slope (dimensionless) and intercept (kcal·mol$^{-1}$).

$$\min_{A_L} \sum_i [A_L(\text{BSSE}_{L,i} - \text{BSSE}_{L-1,i}) + \text{BSSE}_{L,i}]^2 \quad (10)$$

the solution for which is easily found to be

$$A_L = \frac{\sum_i \text{BSSE}_{L,i}(\text{BSSE}_{L-1,i} - \text{BSSE}_{L,i})}{\sum_i (\text{BSSE}_{L-1,i} - \text{BSSE}_{L,i})^2} \quad (11)$$

For those who prefer extrapolations in the familiar $L^{-\alpha}$ form or the Petersson "shift" form $(L+\beta)^{-3}$, the exponents and shifts are easily obtained from $A_L$ as follows (e.g., ref 23)

$$\alpha = \frac{\log\left(1 + \frac{1}{A_L}\right)}{\log \frac{L}{L-1}} \quad (12)$$

$$\beta = \frac{1}{\left(1 + \frac{1}{A_L}\right)^{1/3} - 1} + 1 - L \quad (13)$$

## ■ RESULTS AND DISCUSSION

**Initial Exploration with 24 Heavy-Atom Diatomics.** At first, we started out with a sample consisting of the 24 nonhydrogen diatomics in the W4-17 dataset. For these, we were able to carry out calculations through cc-pV(7+d)Z, heavy-aug-cc-pV(7+d)Z, and 7ZaPa with relative ease; we did so using Gaussian 16, as the largest basis sets entail $k$ functions and MOLPRO can handle $i$ functions at most. (Data for subsequent tables were generated using MOLPRO, which (see the Appendix to ref 50) carries out semicanonicalization *after* integral transformation rather than before like most other codes; hence, fitted parameters for the 24-system dataset may differ subtly, on the order of 1−2 units in the third decimal place.)

For energetic comparisons near the one-particle basis set limit, we employed explicitly correlated data obtained through the rigorous CCSD(F12*) method[52] with the aug-cc-pwCV5Z basis set and, in the F12 geminal, an exponent of 1.4. These data were extracted from the Supporting Information of ref 53; in previous work on a smaller sample,[54] aug-cc-pwCV5Z was found to agree to 0.013 kcal·mol$^{-1}$ RMS with the effectively saturated "Reference-$h$" basis set of Hill et al.[55] We believe that a conservative error bar on our reference data would be about twice that, rounded upward, or 0.03 kcal·mol$^{-1}$. Hence any two extrapolations whose error statistics differ by less than that need to be regarded as of equivalent quality.

As one can see in Table 1, while the Schwenke extrapolation parameters obtained through BSSE minimization are slightly different from literature values obtained through other approaches, they largely follow the same trend. Moreover, the difference between the RMSDs of BSSE-minimizing and TAE-error-minimizing extrapolations is within the uncertainty of the reference values.

The remaining BSSE upon extrapolation is still somewhat significant (0.12 kcal·mol$^{-1}$) for {3,4}ZaPa, but dwindles to 0.05 kcal·mol$^{-1}$ for {4,5}ZaPa and to essentially nil beyond that (0.02 and 0.01 kcal·mol$^{-1}$, respectively, for {5,6}ZaPa and {6,7}ZaPa).

We also obtained a different set of parameters by minimizing the RMSD with respect to CCSD(F12*)/awCV5Z for this sample of 24 molecules. Unsurprisingly, this yields the lowest RMSDs of the three parameter sets — but the differences with BSSE-minimizing extrapolation, except possibly for the {4,5}ZaPa basis set pair, are again within the uncertainty of the reference values.

Using the counterpoise, rather than raw, TAEs leads to marginally different Schwenke coefficients, except for the haV{T,Q}Z+d pair where also TAE(BSSE) differs significantly.

For the connected triple excitations, BSSE minimization in the $n$ZaPa series yields parameters fairly similar to those published by Ranasinghe and Petersson.[21] In ref 56, their {6,7}ZaPa extrapolation was found to essentially represent basis set limits: our BSSE minimization has an RMSD of 0.05 kcal·mol$^{-1}$ for the smallest basis set pair considered ({3,4}ZaPa), but for {4,5}ZaPa this already drops down to 0.01 kcal·mol$^{-1}$, and for {5,6}ZaPa to 0.004. The RMSD(TAE) based minimizations lead to 0.02, 0.01, and 0.003 kcal·mol$^{-1}$, hence only for the smallest basis set pair could the difference be considered even remotely significant. For the cc-pV(n+d)Z sequence, {T,Q}, {Q,5}, and {5,6} pairs all have essentially the same errors for the two sets of parameters: only for the {6,7} pair where the two procedures yield Schwenke parameters that differ by 0.3 (!) is there even a 0.01 kcal·mol$^{-1}$ error difference. Its practical relevance is dubious, given that the {5,6} and even {Q,5} basis set pairs yields similar-quality (T) contributions at much lower cost.

Finally, we considered if the old "NASA recipe"[28] of using a coefficient times the negative BSSE as a basis set incompleteness correction has any practical merit. We thus obtained coefficients more similar to 5/2 than to 3/2 — but more importantly, the RMSD are 3−5 times larger than what can be obtained by two-point extrapolation.

**Further Exploration with (Most of) W4-17 for the CCSD Correlation Energy.** The reader might object that two dozen diatomics is hardly a representative sample, chemically speaking. Here we repeated our analysis to nearly all of the W4-17 thermochemical benchmark, which is eight times larger.

For the CCSD correlation component, Table 2 presents extrapolation parameters, RMS(BSSE) (root-mean-square BSSE), and RMSD(TAE) (root-mean-square deviations in the total atomization energy) for several basis set sequences. (For the largest basis sets, a handful of species had to be omitted for reasons of resource constraints or, in the case of benzene, near-linear dependence of the basis set.) Once again CCSD(F12*)/awCV5Z correlation energies extracted from the ESI of ref 53 were used as the reference.

For the *cc-pV(n+d)Z* family, the agreement between BSSE-minimizing and RMSD(TAE)-minimizing Schwenke parameters can only be described as remarkable: fitted to the W4-08 subset, we have 0.717 vs 0.690 for V{T,Q}Z+d, 0.879 vs 0.883 for V{Q,5}Z+d, and 1.063 vs 1.094 for V{5,6}Z. These differences are well within overlapping 2$\sigma$ uncertainties on the fitted linear regression parameters. The RMSDs in both BSSEs and TAEs are statistically equivalent between the two basis set sequences. Only for {6,7} (Table 1) do we find a significant discrepancy of 1.794 vs 1.400: the RMSD(TAE) if we substitute the former Schwenke parameter for the latter rises from 0.04 to 0.1 kcal·mol$^{-1}$ — still, not much larger than the estimated uncertainty in the reference values. RMS(BSSE) values obtained with the two extrapolation parameters are not appreciably different.

It is well-known (and standard practice in high-accuracy thermochemistry protocols like W4 theory[50,57] and HEAT[58−61]) that adding diffuse functions speeds up basis set convergence especially if highly electronegative elements like O and F are involved. For the haVnZ+d sequence, the Schwenke parameters obtained by BSSE(CBS) minimization and by RMSD(TAE) minimization are again quite similar for the haV{Q,5}Z+d and haV{5,6}Z+d basis set pairs, the resulting RMS(BSSE) and RMSD(TAE) values being statistically equivalent. There is, however, a more pronounced difference for haV{T,Q}Z+d, $A_L$ = 0.782 vs 0.603 when fitted to the W4-08 subset. The BSSE-minimizing $A_L$ = 0.782 yields a quite poor RMSD(TAE) = 0.83 kcal·mol$^{-1}$, almost three times the value obtained with $A_L$ = 0.603.

The similarity between the $A_L$ values obtained from (most of) W4-17 and of its smaller subsets W4-08 and W4-11 is indicative of the stability of the fits, especially for the smaller basis sets where we were able to include all W4-17 species.

As a further sanity check: instead of adjusting a single scaling factor, we carried out linear regression including an intercept that corresponds to correcting for a putative constant bias in the atomization energies. Ideally, said intercept should be as close to zero as possible. For this check, we used the W4-08 subset throughout as we were able to run all its species for all basis sets through $n$ = 6, and hence we can make a fair comparison between the basis set families. In the BSSE fit, the intercept amounts to 0.10 kcal·mol$^{-1}$ for the haV{T,Q}Z+d pair, but drops to insignificant values of 0.04 and 0.013 kcal·mol$^{-1}$ for {Q,5} and {5,6}, respectively.

When fitted to RMSD(TAE) instead, both {T,Q} and {Q,5} have significant intercepts at −0.13 at 0.09 kcal·mol$^{-1}$, respectively. For nZaPa there is a significant intercept for {T,Q} but not for the larger basis set pairs, and concomitantly with that, the Pearson coefficients of determination $R^2$ for the BSSE fits increase sharply from 0.86 to 0.99 and 0.98, respectively, while the corresponding $R^2$ values for the TAE-fits are 0.93, 0.96, and 0.99, respectively.

Figure 2 presents the median BSSE for the CCSD correlation component of TAE across the W4-11 subset for various basis set sequences. For each of these, the BSSE approximately halves with each step in $n$. (The same is seen for the triples, Figure 3.)

The nZaPa series, for smaller $n$, actually seems slightly more prone to BSSE than heavy-aug-pV(n+d)Z. Replacing haV(T

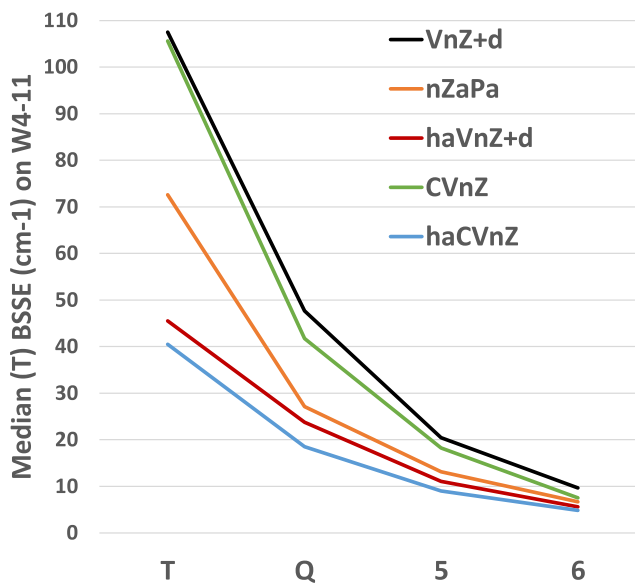

Figure 3. Median BSSE (cm$^{-1}$) for TAE[(T)] over the W4-11 dataset for different basis set sequences.

+d)Z by the *spdf* part of haV(Q+d)Z; haV(Q+d)Z by the *spdfg* part of haV(5+d)Z; and so forth — i.e., the next basis set up with the top angular momentum deleted — drives down the BSSE to the same range as haV(n+1)Z+d.

If (for additional radial flexibility) we apply the cc-pCV$n$Z core–valence basis set sequence to valence correlation, we find that it behaves essentially like the underlying cc-pV$n$Z(+d) series. (There is no need to add tight $d$ functions on second-row elements to a core–valence basis set, as the latter already will include tight $d$ functions to describe especially 2p core–valence correlation.) Only for the cc-pCV{5,6}Z basis set pair is there a semisignificant discrepancy between BSSE- and TAE-based extrapolation coefficients — which in fact goes away when doing linear regression with an intercept. In contrast, for the diffuse function-augmented haCV$n$Z sequence, there is a significant difference (also in RMSD) for the {T,Q} basis set pair: interestingly, here too it disappears when an intercept is allowed into the fit. The said intercept, at +0.1 kcal·mol$^{-1}$ for the BSSE fit and −0.1 kcal·mol$^{-1}$ for the TAE fit, is however a bit large for the authors' comfort. By comparison with the V$n$Z+d and haV$n$Z+d findings, we infer that the "destabilizing" factor here are the diffuse functions.

As shown earlier in ref 41, using the haCV$n$Z core–valence basis sets (Figure 2) for the valence correlation energy does drive down BSSE considerably. Interestingly, combining the *dfg...* functions from haV$n$Z+d with the *sp* set from haV(n+1)Z+d — which combination we denote haV$n$Z+spd — seems to be about equally effective in that regard.

For the {5,6} pair and energy-optimized extrapolations, the differences between the various basis set families are too small to make meaningful distinctions.

**Further Consideration of (T) for a Larger Sample.** It has been shown in great detail (see ref 56 and references therein) that basis set convergence of (T) is considerably faster than for the correlation energy overall; specifically, it was found there that for the W4-08 subset of W4-17, {4,5}ZaPa extrapolation of (T) with the Ranasinghe-Petersson formula[21] causes an RMSD error in TAE[(T)] of just 0.01 kcal·mol$^{-1}$ compared to (T){6,7}ZaPa. Even for the {T,Q} pair this only rose to 0.05

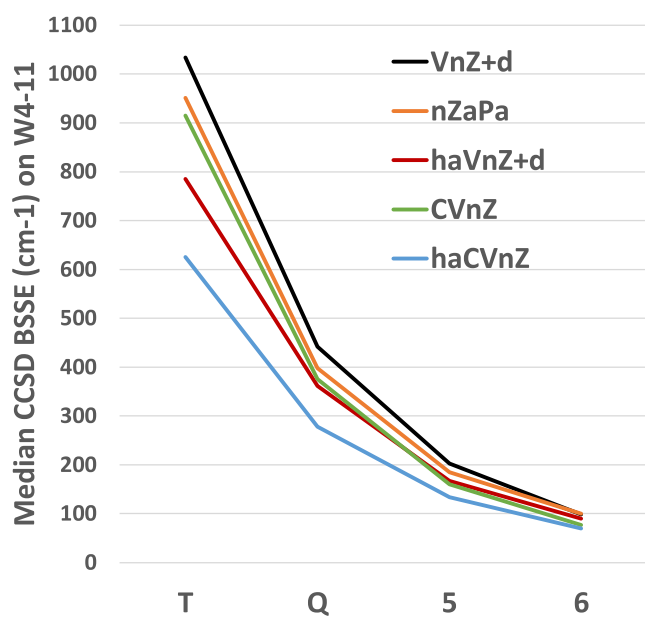

Figure 2. Median BSSE (cm$^{-1}$) for TAE[CCSD$_{corr}$] over the W4-11 dataset for different basis set sequences.

Table 3. Schwenke Extrapolation Coefficients $A_L$ and RMSD Deviations (kcal·mol$^{-1}$) for the Connected Triples Contribution (T) to the Total Atomization Energy

|  |  | BSSE-fitted | | | TAE-fitted | | |
|---|---|---|---|---|---|---|---|
|  |  | TQ | Q5 | 56 | TQ | Q5 | 56 |
| **V$n$Z+d** | $A_L$ W4-11 | 0.730 | 0.754 | 0.985 | 0.721 | 0.804 | 1.085 |
|  | $A_L$ W4-08 | 0.696 | 0.795 | 1.001 | 0.744 | 0.802 | 1.069 |
|  | RMSD BSSE(TAE) W4-11 | 0.049 | 0.013 | 0.010 | 0.049 | 0.013 | 0.007 |
|  | RMSD BSSE(TAE) W4-08 | 0.052 | 0.014 | 0.009 | 0.044 | 0.014 | 0.007 |
| $n$ZaPa | $A_L$ W4-11 | 0.571 | 0.894 | 1.021 | 0.661 | 0.796 | 1.096 |
|  | $A_L$ W4-08 | 0.565 | 0.906 | 1.056 | 0.673 | 0.797 | 1.065 |
|  | RMSD BSSE(TAE) W4-11 | 0.057 | 0.022 | 0.006 | 0.029 | 0.008 | 0.006 |
|  | RMSD BSSE(TAE) W4-08 | 0.060 | 0.022 | 0.005 | 0.029 | 0.009 | 0.005 |
| CV$n$Z | $A_L$ W4-11 | 0.672 | 0.815 | 1.008 | 0.700 | 0.813 | 1.066 |
|  | $A_L$ W4-08 | 0.649 | 0.878 | 1.014 | 0.726 | 0.812 | 1.059 |
|  | RMSD BSSE(TAE) W4-11 | 0.051 | 0.020 | 0.008 | 0.045 | 0.014 | 0.007 |
|  | RMSD BSSE(TAE) W4-08 | 0.057 | 0.021 | 0.008 | 0.041 | 0.016 | 0.007 |
| **haV$n$Z+d** | $A_L$ W4-11 | 0.927 | 0.838 | 1.009 | 0.678 | 0.820 | 1.224 |
|  | $A_L$ W4-08 | 0.838 | 0.899 | 0.989 | 0.706 | 0.798 | 1.190 |
|  | RMSD BSSE(TAE) W4-11 | 0.085 | 0.017 | 0.016 | 0.041 | 0.010 | 0.005 |
|  | RMSD BSSE(TAE) W4-08 | 0.069 | 0.019 | 0.013 | 0.038 | 0.008 | 0.004 |
| haCV$n$Z | $A_L$ W4-11 | 0.754 | 0.969 | 1.152 | 0.643 | 0.830 | 1.208 |
|  | $A_L$ W4-08 | 0.715 | 0.976 | 1.167 | 0.664 | 0.819 | 1.168 |
|  | RMSD BSSE(TAE) W4-11 | 0.044 | 0.023 | 0.005 | 0.032 | 0.007 | 0.005 |
|  | RMSD BSSE(TAE) W4-08 | 0.037 | 0.024 | 0.004 | 0.031 | 0.007 | 0.004 |

kcal·mol$^{-1}$. Hence, we shall eschew overanalysis of results with {Q,5}, let alone {5,6} basis set pairs.

Optimized parameters and performance statistics for the connected triples contribution to TAE can be found in Table 3. Here we used (T)/{5,6}ZaPa or, for the species where available, (T){6,7}ZaPa from the ESI of ref 56 as the reference.

For the pV$n$Z+d and CV$n$Z basis set sequences, the BSSE-fitted and TAE-fitted Schwenke parameters are again quite similar, as are their statistics. For $n$ZaPa, haV$n$Z+d, and haCV$n$Z there is a more pronounced difference for the smaller basis set pairs; comparison for the {5,6} pair is a somewhat inane exercise, as all sets of parameters except haV$n$Z+d have RMSDs of 0.01 kcal·mol$^{-1}$ or below for the (T) component.

## CONCLUSIONS

In response to our research question — whether basis set extrapolation can viably be obtained from the condition that the extrapolated basis set superposition error should approach zero — we can conclude the following:

1. For cc-pV(n+d)Z basis sets, fitting to reference TAEs or fitting to minimize extrapolated BSSE yields similar extrapolation parameters for {T,Q}, {Q,5} and {5,6} basis set pairs.
2. For other basis set sequences, this is consistently the case for the {Q,5} pair.
3. For the haV{T,Q}Z+d and haCV{T,Q}Z+d pairs there appears to be a basis set imbalance in terms of BSSE. This is much less the case for {3,4}ZaPa.
4. For 5Z and 6Z basis sets, the two approaches may still lead to different extrapolation parameters. However, owing to the smaller basis set incompleteness, the predicted basis set limits are of comparable quality considering the uncertainty in the reference values.
5. This recipe becomes less workable for angular momenta beyond $i$ functions, as the BSSEs become too small to form a reliable foundation for fitting.

Thus, basis set extrapolation can be rationalized through BSSE minimization, which eliminates the need to rely on either analytical archetypes about the partial-wave or principal expansions, or on fitting against any sort of external reference energetics. Moreover, since no explicit connection with either the partial-wave or principal expansions exists, it may be applicable to basis set sequences that are not tied to increasing $L$.

## ■ ASSOCIATED CONTENT

### Ⓢ Supporting Information

The Supporting Information is available free of charge at https://pubs.acs.org/doi/10.1021/acs.jpca.4c03012.

> Microsoft Excel workbook containing the raw data for Table 1 as well as its processing; Microsoft Excel workbook containing the raw data for Tables 2−3 and Figures 2−3, as well as its processing (ZIP)

## ■ AUTHOR INFORMATION

### Corresponding Author

**Jan M. L. Martin** − *Department of Molecular Chemistry and Materials Science, Weizmann Institute of Science, 7610001 Reḥovot, Israel; On sabbatical at Quantum Theory Project, University of Florida, Gainesville, Florida 32611, United States;* ◯ orcid.org/0000-0002-0005-5074; Phone: +972 8 9342533; Email: gershom@weizmann.ac.il; Fax: +972 8 9343029

### Authors

**Vladimir Fishman** − *Department of Molecular Chemistry and Materials Science, Weizmann Institute of Science, 7610001 Reḥovot, Israel;* ◯ orcid.org/0009-0004-7570-136X

**Emmanouil Semidalas** − *Department of Molecular Chemistry and Materials Science, Weizmann Institute of Science, 7610001 Reḥovot, Israel;* ◯ orcid.org/0000-0002-4464-4057

Complete contact information is available at:
https://pubs.acs.org/10.1021/acs.jpca.4c03012

### Notes

The authors declare no competing financial interest.

## ACKNOWLEDGMENTS

This work was supported by the Israel Science Foundation (grant 1969/20) and by the Uriel Arnon Memorial Center for AI research into smart materials. J.M.L.M. thanks the Quantum Theory Project at the University of Florida and its head, Prof. John F. Stanton, for their hospitality, and Dr. Nisha Mehta for discussions on BSSE in chalcogen bonding that provided some inspiration for the present work. E.S. acknowledges doctoral scholarships from the Feinberg Graduate School (Weizmann Institute of Science) and the Onassis Foundation (Scholarship ID: FZP 052-2/2021-2022). All calculations were carried out on the ChemFarm HPC cluster of the Weizmann Institute Faculty of Chemistry.